\documentclass[12pt,preprint]{aastex} 

\IfFileExists{srcltx.sty}{\usepackage[active]{srcltx}} 

\slugcomment{Submitted to the {\it Astrophysical Journal}}
\slugcomment{UCLA/09/TEP/57}
\shorttitle{Sterile Neutrinos in Willman 1}
\shortauthors{Loewenstein and Kusenko}

\begin{document}
\title{Dark Matter Search Using {\em Chandra} Observations of Willman
1, and a Spectral Feature Consistent with a Decay Line of a 5~keV
Sterile Neutrino}

\author{Michael Loewenstein\altaffilmark{1,2}, Alexander
Kusenko\altaffilmark{3,4}}

\altaffiltext{1}{Department of Astronomy, University of Maryland,
College Park, MD.}
\altaffiltext{2}{CRESST and X-ray Astrophysics Laboratory NASA/GSFC,
Greenbelt, MD.}
\altaffiltext{3}{Department of Physics and Astronomy, University of
California, Los Angeles, CA 90095-1547, USA}
\altaffiltext{4}{Institute for the Physics and Mathematics of the Universe,
University of Tokyo, Kashiwa, Chiba 277-8568, Japan}


\begin{abstract}
We report the results of a search for an emission line from
radiatively decaying dark matter in the {\it Chandra} X-ray
Observatory spectrum of the ultra-faint dwarf spheroidal galaxy
Willman 1. 99\% confidence line flux upper limits over the 0.4-7 keV
{\it Chandra} bandpass are derived and mapped to an allowed region in
the sterile neutrino mass-mixing angle plane that is consistent with
recent constraints from {\it Suzaku} X-ray Observatory and {\it
  Chandra} observations of the Ursa Minor and Draco dwarf
spheroidals. A significant excess to the continuum, detected by
fitting the particle-background-subtracted source spectrum, indicates
the presence of a narrow emission feature with energy $2.51\pm 0.07
(0.11)$ keV and flux $[3.53\pm 1.95 (2.77)]\times 10^{-6}\ {\rm
  photons}\ {\rm cm}^{-2}\ {\rm s}^{-1}$ at 68\% (90\%)
confidence. Interpreting this as an emission line from sterile
neutrino radiative decay, we derive the corresponding allowed range of
sterile neutrino mass and mixing angle using two approaches. The first
assumes that dark matter is solely composed of sterile neutrinos, and
the second relaxes that requirement. The feature is consistent with
the sterile neutrino mass of $5.0\pm 0.2 $~keV and a mixing angle in a
narrow range for which neutrino oscillations can produce all of the
dark matter and for which sterile neutrino emission from the cooling
neutron stars can explain pulsar kicks, thus bolstering both the
statistical and physical significance of our measurement.
\end{abstract}



\section{Introduction}

\subsection{Context}

None of the Standard Model particles can account for the dark matter
that makes up most of the mass in the universe. We report the latest
results of a search for dark matter in the form of relic sterile
neutrinos (see Kusenko 2009 for an up-to-date review). The motivation
for considering the sterile neutrino, one of a number of feasible
dark-matter candidates, is two-fold. First, the discovery of ordinary
neutrino masses is most easily accommodated by means of the so-called
seesaw mechanism, which calls for some new gauge-singlet fermions. If
all of these fermions have very large Majorana masses, there are no
additional degrees of freedom (particles) at the low energy
scale. However, if one of these mass parameters lies in the 1--30~keV
range, the corresponding {\em sterile} neutrino can be the long
sought-after dark matter particle~\citep{dw94}. Second, the physics of
supernovae, that often assists in ruling out hypothetical low-mass
particles, provides some intriguing clues in favor of the existence of
sterile neutrinos with the same parameters that are required to
explain dark matter. Such particles would be anisotropically emitted
from a cooling newly born neutron star, inducing a sufficient recoil
momentum in the neutron star to explain the observed velocities of
pulsars~\citep{ks97,fkmp03}. X-ray observations offer the best probe
of this well-motivated dark-matter candidate~\citep{K09}. In the early
universe, sterile neutrinos may be produced through non-resonant
oscillations \citep{dw94}, resonant oscillations~\citep{sf99} and via
various other channels~\citep{k06,st06,pk08}. The kinetic properties
of dark matter relevant for structure formation on small scales (i.e.,
how ``warm'' or ``cold'' the dark matter is) depend on the production
scenario, as well as the particle mass~\citep{k06,p08,boyan08b}.

The generic prediction is that relic sterile neutrinos must decay into
a lighter neutrino and a photon~\citep{pw82}. Since this two-body
decay of a 1--30~keV mass particle produces a narrow line with energy
$E_\gamma = m_{\rm st}c^2/2$, X-ray astronomy provides a unique
opportunity to discover these relic particles. Archival X-ray data has
been used to set limits on relic sterile neutrinos (see the review in
Kusenko 2009). In addition, a dedicated search for sterile neutrinos
using the {\em Suzaku} X-ray telescope has recently been conducted
by~\cite{lkb09} (hereafter, LKB). Here we will present new results
from the search using the {\em Chandra} X-ray Observatory. Our
analysis of data on the Willman 1 dwarf spheroidal galaxy yields
evidence of a 2.5~keV line, consistent with the decay of a relic
sterile neutrino with mass 5~keV; and, with an inferred line strength
consistent with that expected if all of the dark matter is composed of
sterile neutrinos produced by oscillations.

\subsection{Target selection}

Given the uncertainty in sterile neutrino properties, and the
importance of dark matter discovery, one ought to explore the X-ray
band with existing instruments using well-chosen targets. While the
decay timescale greatly exceeds the Hubble time~\citep{pw82,barger},
the emissivity may reach detectable levels in regions of large dark
matter surface density, such as galaxy clusters, M31, and Local Group
dwarf spheroidal galaxies~\citep{afp01,aft01,dh02,bnrst06}. The last
offer, arguably, the most promising opportunity because of the
proximity and high dark matter density in these systems, and the
absence of additional competing internal X-ray sources. {\it Suzaku}
Observing Cycle 2 observing time was awarded time to study the best
available targets at that time: the Ursa Minor and Draco dwarf
spheroidals (LKB). Subsequently, a new sub-population of faint Milky
Way satellites discovered in the Sloan Digital Sky Survey (SDSS) were
found to have very high mass-to-light ratios
\citep{str08a,wal09}. These rival or surpass Ursa Minor and Draco as
compelling targets for sterile neutrino searches. We were awarded 100
ks of observing time in {\em Chandra} cycle 10 to investigate one of
these -- namely Willman 1. These data represent the first X-ray
observations of this extreme system; and, we report on our findings
here.

\subsection{Previous Results on Ursa Minor}

The basic equations relating the galaxy parameters (distance
100$d_{100}$ kpc, dark matter mass in projection $M_{\rm
  pro}=10^{7}M_{7}$ M$_{\odot}$), X-ray observables (line energy
$E_\gamma$, line flux $F_{\rm line}$), and sterile neutrino parameters
(mass $m_{\rm st}$, mixing angle $\theta$, fraction of dark matter in
sterile neutrinos $f_{\rm st}$), are the following (Loewenstein et
al. 2009, Kusenko 2009, and references therein):
\begin{equation}
m_{\rm st}=2E_\gamma,
\end{equation}
\begin{equation}
\Gamma_{\nu_s\rightarrow\gamma \nu_a}=5.52\times 10^{-32}\left(
\frac{\sin^2 \theta}{10^{-10}} \right) \ \left( \frac{m_{\rm st}}{\rm
keV}\right)^5,
\end{equation}
and
\begin{eqnarray}
F_{\rm line}& =& 4.69\times 10^{-6}~\Gamma_{-27} \, f_{\rm st} \,
M_{7}\, d_{100}^{-2} \left(\frac{\rm keV}{E_\gamma} \right) \\ & = &
5.15\times 10^{-10} \left(\frac{\sin^2 \theta}{10^{-10}} \right)
\left(\frac{m_{\rm st}}{\rm keV} \right)^4\, f_{\rm st}\, M_{7}\,
d_{100}^{-2} \ {\rm photons} \ {\rm cm}^{-2} \ {\rm s}^{-1}.
\end{eqnarray}
The abundance of sterile neutrinos produced through non-resonant
oscillations may be calculated for given values of $m_{\rm st}$ and
$\theta$, assuming the standard early thermal history of the
universe. This relation determines $f_{\rm st}$ as a function of
$m_{\rm st}$ and $\theta$, assuming that all sterile neutrinos are
produced in this way as first proposed by \cite{dw94}, and lower
limits on $f_{\rm st}$ if there is some non-negligible lepton
asymmetry~\citep{sf99} or if some other production channels
contributed more than the oscillations to the sterile neutrino relic
abundance~\citep{K09}.

LKB did not detect a sterile neutrino emission line in Ursa Minor {\em
Suzaku} spectra. The line flux upper limit was used to derive an
excluded region in the $m_{\rm st}-\theta$ plane for the case where
all of the dark matter is composed of sterile neutrinos. A more
general constraint was inferred from the condition that the line flux
upper limit not be overproduced through oscillations at the sterile
neutrino relic density corresponding to each pair ($m_{\rm
st},\theta$). We derived line flux upper limits using a
maximum-likelihood approach that, along with the lack of intrinsic
X-ray emission, enabled us to minimize systematics, and account for
those that remained. The resulting constraints on sterile neutrinos
are illustrated in Figure 9 of LKB. Although the constraints from
this single {\it Suzaku} observation are comparable to previous
results over the entire 1--20 keV mass, the allowed range does not
rule out sterile neutrinos as a viable dark matter candidate. We
therefore expanded our search to the Willman 1 system -- conducting
the first X-ray observation of one of the new SDSS dwarf spheroidal
galaxies. The capabilities of the {\it Chandra} X-Ray Observatory are
well-suited to the compact size of Willman 1; and, these data provide
a good complement to the broader beam observations of the more
extended classical Milky Way dwarf spheroidals (Section 4.2).

\section{Data Analysis}

\subsection{Observation and Data Processing}

Willman 1 ({\it Chandra} ObsID 10534, SeqNum 600727) was observed by
{\it Chandra} on January 27-28 2009 for 102745.5 seconds. The data was
taken in TIMED VFAINT mode, and was pipeline-processed on 2009-2-10
using ascdsver 7.6.11.10. We use CIAO version 4.1.2 (CALDB 4.1.4), and
XSPEC version 12.5 in our analysis. Chips 0, 1, 2, 3, 6, 7 were on
during the observation. Willman 1 was positioned at the ACIS-I
aimpoint, and we generally restrict our analysis to the I-array (chips
0-3) data.

We analyze the standard level 2 event file, and also reprocess the
level 1 event file in order to implement the VFAINT data-mode
quiescent background reduction
technique.\footnote{http://cxc.harvard.edu/ciao/threads/aciscleanvf} For
the latter, likely background events are flagged based on the pulse
heights in border pixels of the 5x5 pixel event islands; default split
threshold and trail parameters are adopted. These events, as well as
those corresponding to bad grades, are removed as the new level 2
event file is generated. No flares are found upon examination of the
light curves; and, time intervals where the count rate deviates by
more than 10\% of the mean are excised. The final exposure times are
99400 and 97100 seconds for the cleaned pipeline processed and
reprocessed event files, respectively.

\subsection{The Background}

We consider off-axis (S2, {\it i.e.} chip 6), blank sky, and stowed
data sets as candidates for constructing background spectra. The last
corresponds to events collected with the ACIS stowed while the HRC-I
is in the focal plane; and, has been demonstrated to be consistent
with the dark
moon.\footnote{http://cxc.harvard.edu/contrib/maxim/stowed/} That is,
it can provide an estimate of the pure particle background (PB). In
contrast, the blank sky and S2 data include the cosmic (CXB) and
galactic (GXB) X-ray backgrounds (which vary over the sky, especially
the GXB). The blank sky and stowed event files are reprojected using
the Willman 1 aspect solution. The quiescent PB reduction procedure
may also be applied to these background event files. If dark matter is
composed of sterile neutrinos, their presence in the Milky Way results
in line emission in the S2 and blank sky data that will reduce the
signal if subtracted out as background. Additionally, the S2 chip is
still within the solid angle subtended by the Willman 1 dark halo (see
below). Therefore, our constraints are derived from the PB-subtracted
spectra from which no astrophysical sources of extended X-ray emission
have been eliminated.

\subsection{Spectral Analysis Setup}

Source and background spectra are extracted from identical $5'$
circular apertures centered on the Willman 1 position or its
equivalent in detector coordinates, except for the S2 chip background
spectrum that is extracted from a $3'$ circle optimized in a way that
accounts for chip uniformity and point sources near the edge of the
chip. The S2 chip center is $15'$ (165 pc) from the position of
Willman 1, and lies outside the region where stellar velocities are
measured. The allowed dark matter profile thus spans a large range:
the average surface mass density in this field may be as large as
$\sim 50$\%, or as small as $\sim 10$\%, that in the central I-array
extraction region. Regions corresponding to point sources detected in
the Willman 1 image via application of the CIAO wavelet point source
detection algorithm are excluded. Background spectra are re-scaled
based on the relative count rates in the 9-12 keV band where virtually
all events are due to the PB. The PB has been found to vary
significantly in normalization, but not in shape, so that the rescaled
PB spectra can be used to construct an appropriate PB-subtracted
source spectrum.\footnote{http://cxc.harvard.edu/contrib/maxim/bg/}

\subsection{Spectral Analysis}

In our analysis of the {\it Suzaku} data on the Ursa Minor dwarf
spheroidal galaxy, we based our constraints on the total
(unsubtracted) spectrum. Because the {\it Chandra} PB is higher, but
more precisely determined, we adopt a more conventional approach here
and consider background-subtracted spectra, grouped to a minimum of 15
counts per bin, utilizing $\chi^2$ statistics to derive best fits and
parameter uncertainties.

The cleanest PB subtraction is realized utilizing the source and
stowed background spectra with application of the VFAINT data-mode
quiescent background reduction technique, although using the
corresponding pipeline-processed datasets yielded similar results
(with slightly larger uncertainties). The source spectrum is fitted
with a model consisting of a solar abundance thermal plasma (to model
the GXB) and power-law (to model the CXB) absorbed by the Galactic
column density. In this case the GXB component is not required,
although the upper limit on the flux is consistent with the
larger-scale average GXB surface brightness measured in the {\it ROSAT
  All-Sky Survey}. The best-fit ($\chi^2/\nu=1.09$) power-law photon
index is 1.8 (1.4-2.7 at 90\% confidence), and the normalization
consistent with the fraction of the CXB unresolved by our observation
($\sim 25$\%; Mushotzky et al. 2000). Figures 1a and 1b show the
total, (stowed) background, and background-subtracted spectra for
cases utilizing pipeline-processed, and reprocessed
(reduced-background) event lists. The best-fit power-law model, and
fit residuals, are also shown. We show the full 0.3-12 keV energy
band, although the fits are conducted over 0.4-7 keV. There are
PB-related residuals in the background-subtracted spectrum above 7 keV
where the {\it Chandra} sensitivity experiences a steep decline. Some
of these persistent residuals may be attributed to the fact that the
time-dependent gain correction cannot be re-applied to the background
data, as was done with the source data.

\begin{figure}
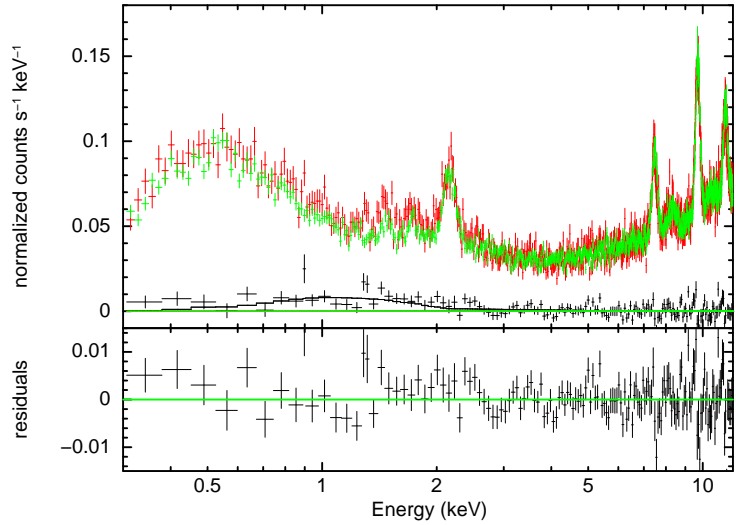

\centering
\includegraphics[scale=0.4,angle=-90]{nfig1a.eps}
\hfil
\caption{{\bf (a) top:} Total (red), PB (green), and source ({\it
i.e.} PB-subtracted; black) spectra. The histogram is the best-fit
GXB+CXB model to the source spectrum. {\bf (b) bottom:} Same as {\bf
top} with the quiescent-particle-background reduced in the total and
PB spectra as explained in the text.}
\hfil 
\centering
\includegraphics[scale=0.4,angle=-90]{nfig1b.eps}
\end{figure}

The blank-sky background (SB) is significantly brighter than the
source spectrum below 1 keV. The total PB-subtracted 0.4-7 keV surface
brightness is $\approx 1.7\times$ higher in the SB than in Willman 1.
Evidently the GXB is brighter in the former and, as a result, the
SB-subtracted spectrum ought to be utilized with care (Figure
2). Also, as previously mentioned, if there is dark matter decay in
the X-ray band there will be a contribution from the Milky Way halo to
the SB that is absent in the PB.

\begin{figure}
\centering \includegraphics[scale=0.6,angle=-90]{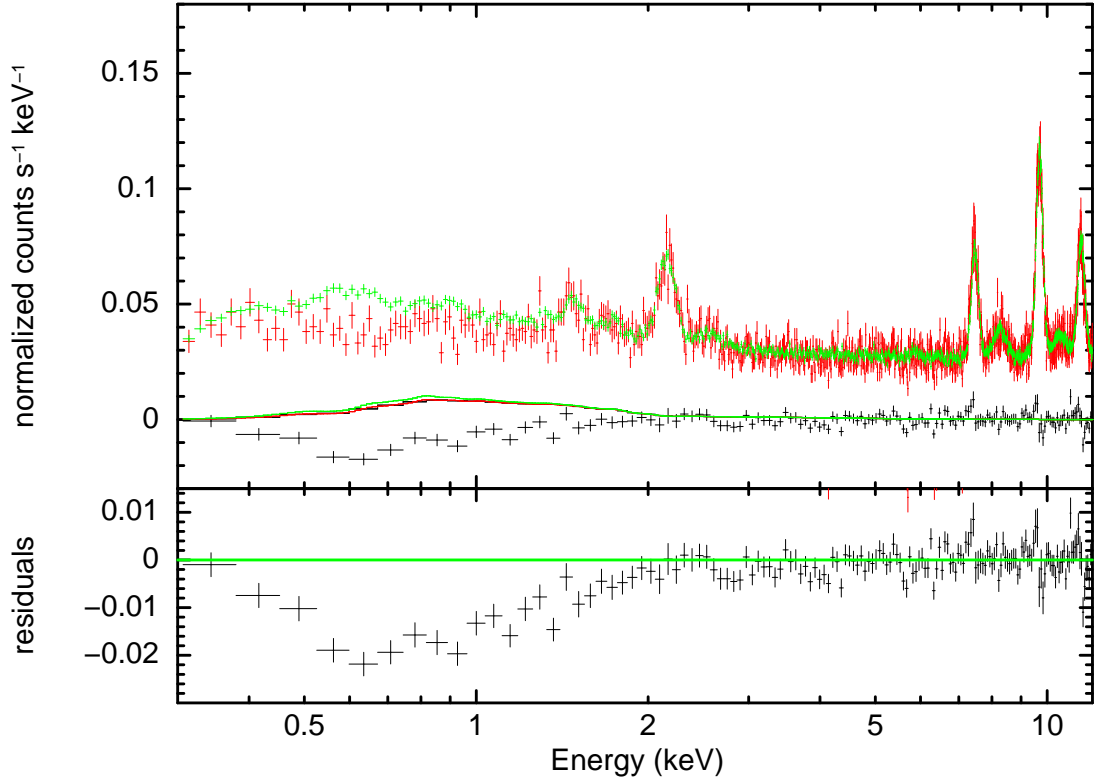} \hfil
\caption{Total (red), SB (green), and SB-subtracted (black)
  spectra. The histogram is the best-fit GXB+CXB model to the {\it
    PB-subtracted} spectrum of Figure 1b. The SB results in an
  oversubtraction below 2 keV. Due to the contribution of the Milky
  Way halo the SB is expected to contain the dark-matter signal as
  well; and, therefore subtraction of the SB can simultaneously lower
  both the background and the signal. The oversubtraction below 2 keV
  and the weakening of the signal due to subtraction of the Milky Way
  contribution are the two reasons why we do not use the SB-subtracted
  spectrum for data analysis.}
\end{figure}

\section{Limits on Sterile Neutrino Parameters from the {\it Chandra}
Spectrum of Willman 1}

In deriving upper limits, we divide the spectrum into low ($<1.1$ keV)
and high ($>1.1$ keV) energy segments, since any GXB emission will be
negligible in the latter. The high energy segment best-fit
($\chi^2/\nu=1.11$) power-law slope is 1.77 (1.22-2.37 at 90\%,
confidence). At low energies, power-law ($\chi^2/\nu=0.95$, slope =
$2.27_{+1.39}^{-1.09}$) and thermal {\bf apec} ($\chi^2/\nu=0.92$,
$kT=0.48\pm0.15$ keV) models provide acceptable fits. We conduct Monte
Carlo numerical experiments (LKB, and references therein) where we
find that, in fits to simulated continua spectra with models that
include an emission line, the fit is improved by $\Delta\chi^2=9.2$
with $\sim 1$\% frequency -- closely approximating that for a $\chi^2$
distribution with two degrees of freedom. We thus adopt
$\Delta\chi^2=9.2$ flux limits, derived from adding an unresolved
Gaussian component in 10 eV intervals over the 0.4-7 keV bandpass to
simulated spectra of the best-fit line-free model. The errors are
averages over 100 simulations that are initiated using parameters from
fits to a seed simulation (and not the best-fit parameters to the
actual data), so that the uncertainties in the parameter values are
taken into account. Line fluxes are allowed to go negative (as is the
case with the simulations). The upper and lower line flux limits, in
200 eV bins, based on the PB-subtracted fits are shown in Figure
3. The latter are negative over the entire 0.4-7 keV bandpass. The
constraints based on the SB-subtracted fits are similar in magnitude
above 2 keV.

\begin{figure}
\centering
\includegraphics[scale=0.8,angle=0]{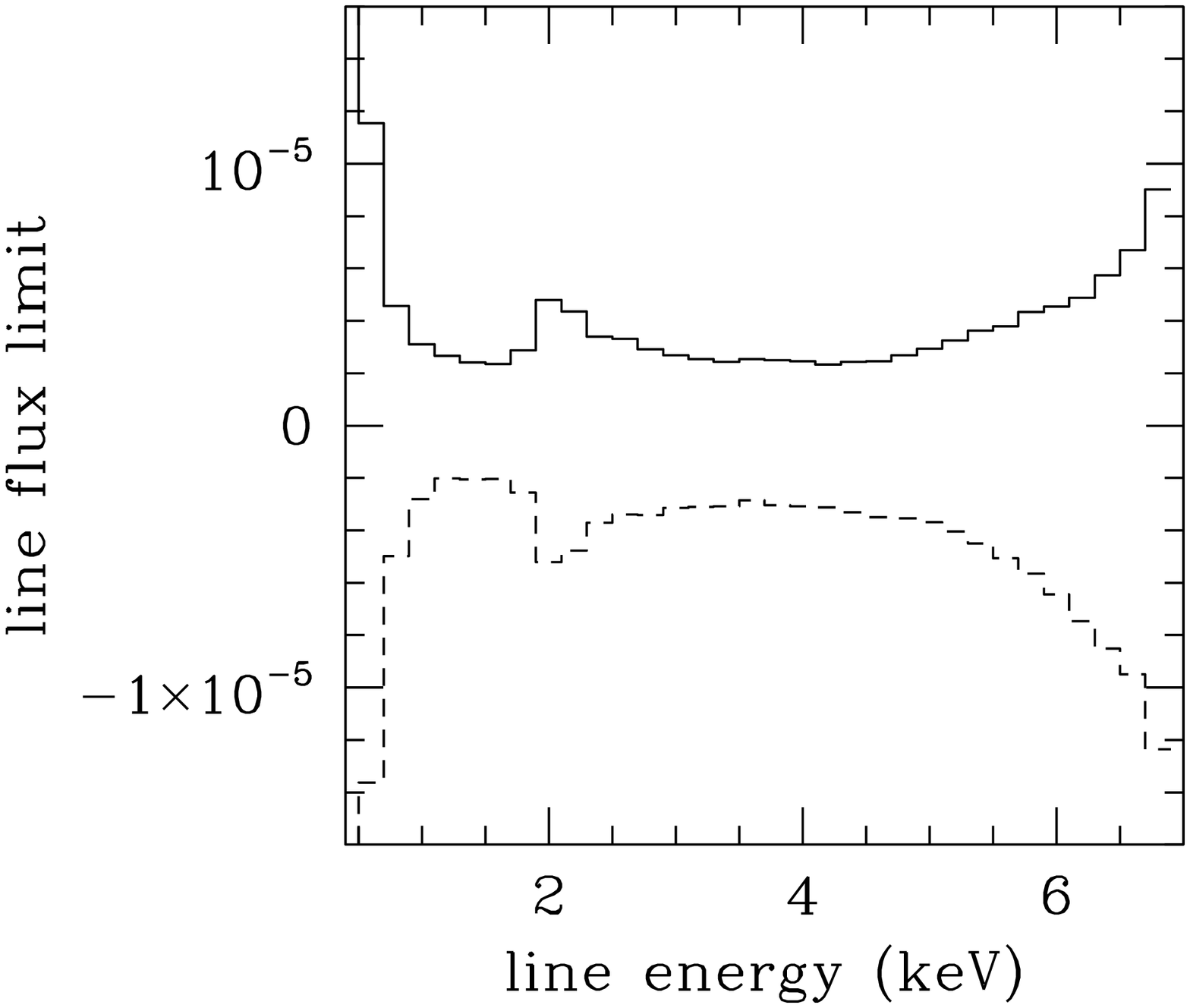}
\hfil
\caption{$\Delta\chi^2=9.2$ line flux upper (solid histogram) and lower
(broken histogram) limits in photons cm$^{-2}$ s$^{-1}$.}
\end{figure}

The mapping of the Willman 1 constraints onto the
sterile-neutrino-mass-mixing-angle plane (equation 4) is shown in
Figure 4a. As discussed above (Section 1.3; see also Section 3 in LKB
and Section 5 below) there are two kinds of limits, both of which are
shown. The first assumes only that the thermal history of the universe
is given by standard Big Bang cosmology. Sterile neutrinos are ruled
out for the region in parameter space where the flux from sterile
neutrinos produced at minimal abundance via the DW mechanism
\citep{asl07} exceeds our derived limits. The second limit is that
which overpredicts the derived limits if sterile neutrinos account for
100\% of the dark matter (dark-matter fraction of sterile neutrinos
$f_{\rm st}=1$), regardless of how they are produced. In Figure 4b,
the former limits, which are the most general, are compared with those
we derived in Ursa Minor with {\it Suzaku}.

The smaller {\it Chandra} beam and more compact Willman 1 dark matter
distribution compensate somewhat for the higher sensitivity of the
{\it Suzaku} measurement to yield constraints at sterile neutrino
masses $<14$ keV (corresponding to photon energies in the spectra $<7$
keV) of similar magnitude to those derived for Ursa Minor in LKB.
This comparison utilizes a projected mass enclosed within the relevant
radius of 55 pc ($5'$ at 38 kpc) $M_{\rm pro}=2\times 10^6$
M$_{\odot}$ -- corresponding to a mass surface density of $\Sigma=210$
M$_{\odot}{\rm pc}^{-2}$ -- as determined by projecting the best-fit
NFW \citep{nfw97} mass model in \cite{str08b}. The 90\% confidence
contours for the NFW parameters allow values of $M_{\rm pro}$ that are
$4\times$ lower, and $2-3\times$ higher. In addition, we estimate that
the Milky Way dark matter halo and the extragalactic dark matter
contribute an additional $65-125$ M$_{\odot}{\rm pc}^{-2}$ (Xue at
al. 2008, and references therein). While maintaining a fiducial mass
of $2\times 10^6$ M$_{\odot}$, we show the effect on the general
constraints of varying the surface density, and hence the effective
mass in the FOV, by factor of two in either direction in Figure 4b.

\begin{figure}
\centering
\includegraphics[scale=0.45]{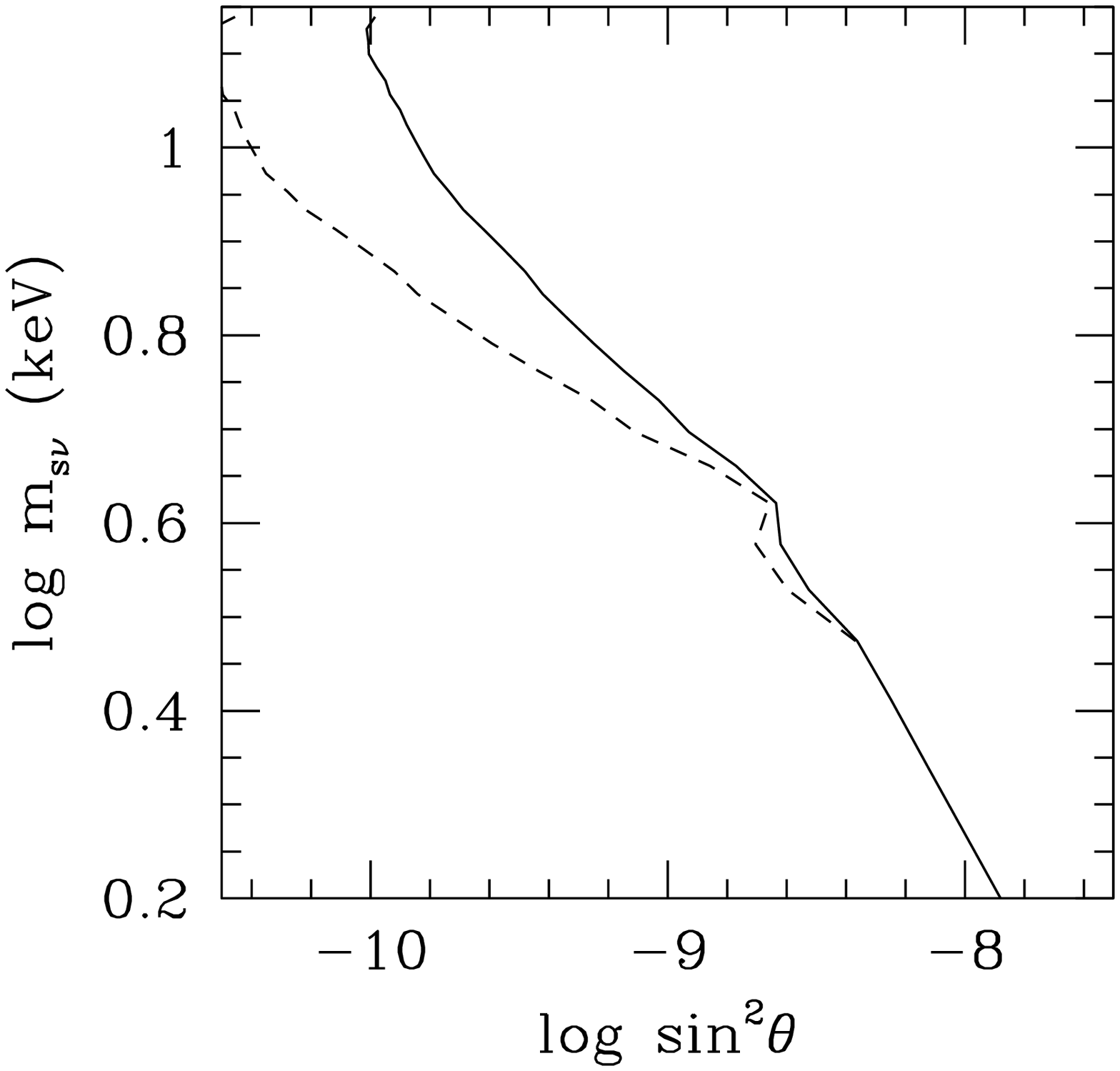}
\hfil
\caption{{\bf (a) top:} The sterile neutrino parameter space to the
  right of the solid curve is excluded by the {\it Chandra}
  observation of Willman 1 assuming only that DW production by
  neutrino oscillations takes place. The sterile neutrino parameter
  space to the right of the broken curve is excluded if 100\% of dark
  matter is composed of sterile neutrinos produced by some
  (unspecified) mechanism. The mass of $2\times 10^6$ M$_{\odot}$ is
  derived from the best-fit NFW \citep{nfw97} mass model in
  \cite{str08b}. {\bf (b) bottom:} The solid curve from (a) is
  reproduced and compared to constraints assuming projected masses of
  1 and $4\times 10^6$ M$_{\odot}$ (broken curves; see text), and to
  the constraints derived from {\it Suzaku} observation of Ursa Minor
  (dotted line).}  \hfil \centering
\includegraphics[scale=0.45]{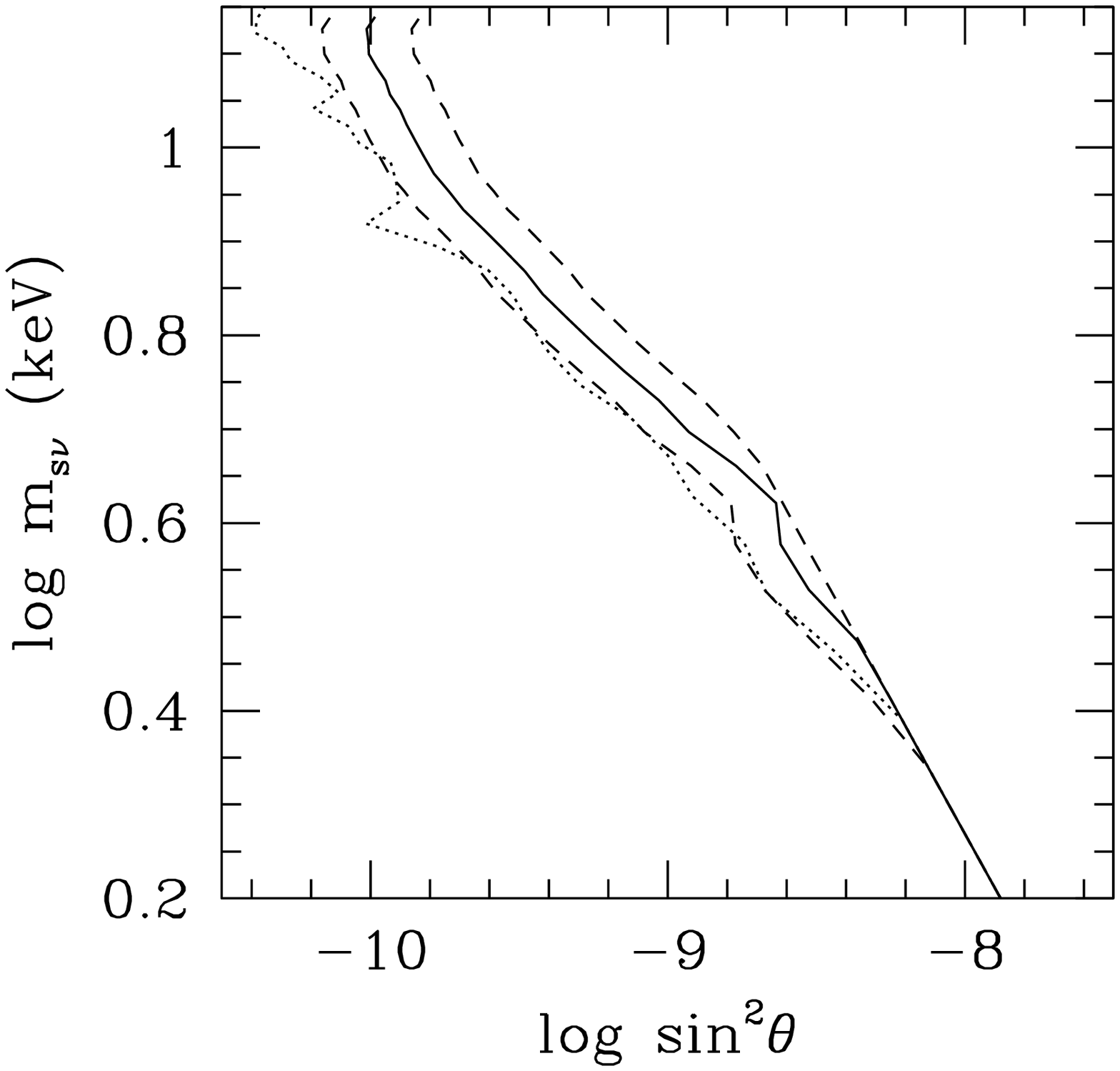}
\end{figure}

\section{A Candidate Line at 2.5 keV}

In addition to the limits derived above, we present evidence of a
spectral feature consistent with a 2.5~keV line from decay of a 5~keV
relic sterile neutrino.

\subsection{Line Flux and Statistical Significance}

The largest residual excess emission to the PB-subtracted Willman 1
spectrum with respect to the best-fit continuum model occurs at $\sim
2.5$ keV. When we add an additional Gaussian component, we find that
$\chi^2$ improves by 5 (Figures 5 and 6). The line-width is consistent
with an unresolved feature broadened only by the instrumental line
spread function, as expected considering the small Willman 1 velocity
dispersion. For power-law-plus-Gaussian fits to the 2.2-5 keV
spectrum, the best-fits and uncertainties on the line energy and flux
are $2.51\pm 0.07 (0.11)$ keV and $[3.53\pm 1.95 (2.77)]\times
10^{-6}\ {\rm photons}\ {\rm cm}^{-2}\ {\rm s}^{-1}$, where the errors
are 68\% (90\%) confidence limits for two degrees of freedom
corresponding to $\Delta\chi^2=2.3$ (4.61) (Section 3; Yaqoob
1998). The corresponding confidence contours are shown in Figure 7, as
is the $\Delta\chi^2=9.21$ (99\% confidence) contour. The 2.2 keV
limit was chosen to avoid the energy associated of the peak of the
brightest of the PB Au fluorescence lines, and with the deep {\it
  Chandra} mirror iridium edge (Chartas et al. 2000; there is also a
shallow edge at $\sim 2.55$ keV that does not impact our results). The
issue of these instrumental features is specifically addressed in
Section 4.2.4. However, extending the range to the full 1.1-7 keV high
energy segment bandpass yields a consistent measurement ($F_{\rm
  line}=[2.85\pm 1.85 (2.57)]\times 10^{-6}\ {\rm photons}\ {\rm
  cm}^{-2}\ {\rm s}^{-1}$) as do fits over other subsets of this
bandpass.  In the 2.4-2.6 keV energy interval, the total, PB, and
source counts are $696\pm 26.4$, $601\pm 13.6$, and $95\pm 30$,
respectively. In the composite models the emission line component
accounts for 90\% of the flux. 20 counts in this interval are expected
based on the best-fit power-law model {\it without a line}. Figure 8
shows the unfolded spectrum -- the ratio of the data values to the
product of the model multiplied by the response, all multiplied by the
model values -- over the 1.5-5 keV spectral region.

\begin{figure}
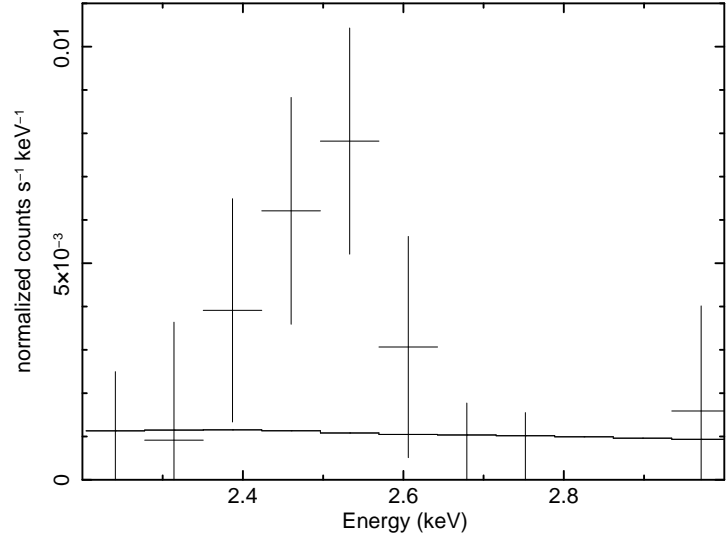

\centering 
\includegraphics[scale=0.4,angle=-90]{nfig5a.eps} 
\hfil 
\caption{{\bf (a)} Data and best-fit power-law model to the 2.2-5 keV
spectrum, in the 2.2-3 keV energy region ({\bf top}), and {\bf (b)}
contributions to $\chi^2$ over 2.2-5 keV({\bf bottom}).}
\hfil 
\centering
\includegraphics[scale=0.4,angle=-90]{nfig5b.eps}
\end{figure}

\begin{figure}
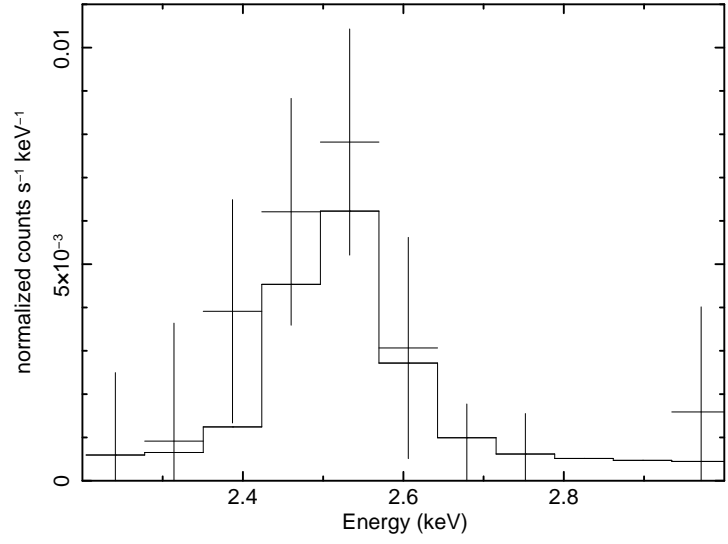

\centering
\includegraphics[scale=0.4,angle=-90]{nfig6a.eps}
\hfil
\caption{Same as Figure 5 for the best-fit power-law-plus-Gaussian
model.}
\hfil 
\centering
\includegraphics[scale=0.4,angle=-90]{nfig6b.eps}
\end{figure}

\begin{figure}
\centering
\includegraphics[scale=0.6,angle=-90]{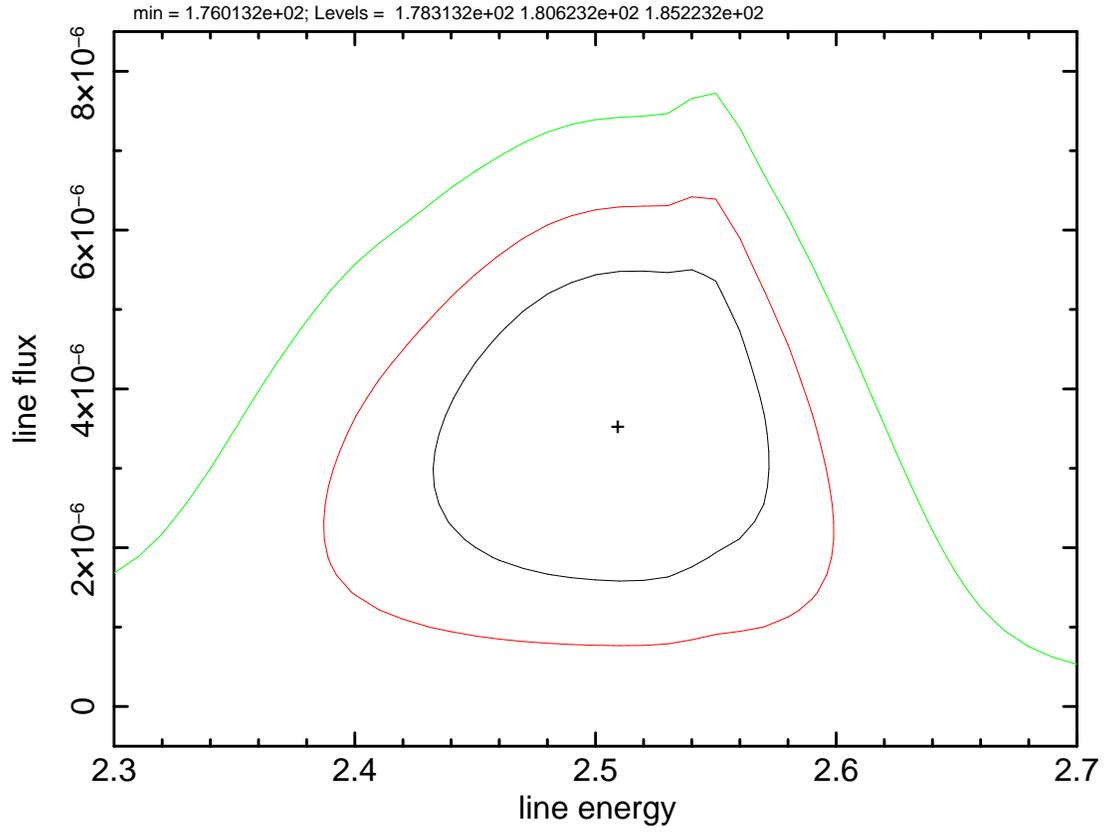}
\hfil
\caption{$\Delta\chi^2=2.3$, 4.61, and 9.21 confidence contours for
the emission line parameters.}
\end{figure}

The feature is not positively detected in the global blank sky
spectrum (Abazajian et al. 2007, and references therein). The
$\Delta\chi^2=4.61$ upper limit based on the reprojected estimated SB
of Willman 1 is $2.1\times 10^{-6}\ {\rm photons}\ {\rm cm}^{-2}\ {\rm
  s}^{-1}$. When we utilize the SB, or appropriately-scaled S2,
spectrum as background the fits are not significantly improved in
models with an additional emission line, although the upper limits are
consistent with the line flux based on the PB-subtracted spectrum. As
the SB (Figure 2) and S2 spectra overestimate the Willman 1
background, and will include a contribution due to the Milky Way dark
matter halo (plus, possibly significant, additional emission from the
outer Willman 1 halo in the latter), the expectation that the
PB-subtracted spectrum is most sensitive for these purposes is
confirmed.

\begin{figure}
\centering
\includegraphics[scale=0.6,angle=-90]{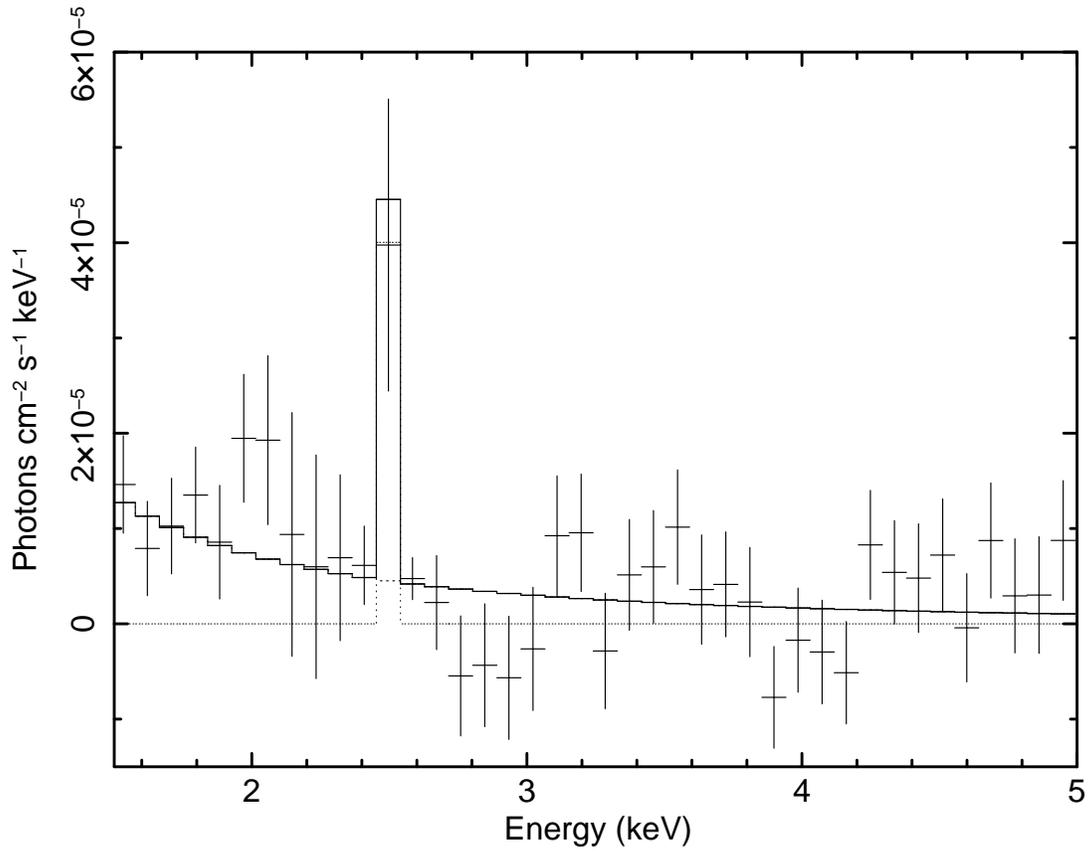}
\hfil
\caption{Unfolded 1.5-5 keV spectrum; see text for details.}
\end{figure}

Notably, the flux is consistent with the narrow region in the
mass-mixing-angle plane where the DW neutrino oscillation mechanism
produces all of the dark matter (and where sterile neutrino emission
from neutron stars may explain pulsar kicks -- see next
section). Figure 9 shows the predicted flux at each energy $E_\gamma$
assuming DW production of sterile neutrinos of $m_{\rm st}=2E_\gamma$,
and $f_{\rm st}=1$. We compare this with the $\Delta\chi^2=4.61$ upper
limits derived from Monte Carlo simulations (see Section 3), and our
measurement. Even accounting for uncertainties in the DW production
rate, and allowing for a factor of four uncertainty in the mass, a
fluctuation at this level maps into this narrow region only over $\sim
1-3.5$ keV. At lower energies the predicted line flux is below our
sensitivity, while at higher energies a much stronger feature would be
expected. That is, the likelihood of a marginal feature compatible
with the simplest sterile neutrino hypothesis is much less than the
likelihood at some random X-ray energy.

\begin{figure}
\centering
\includegraphics[scale=0.8,angle=0]{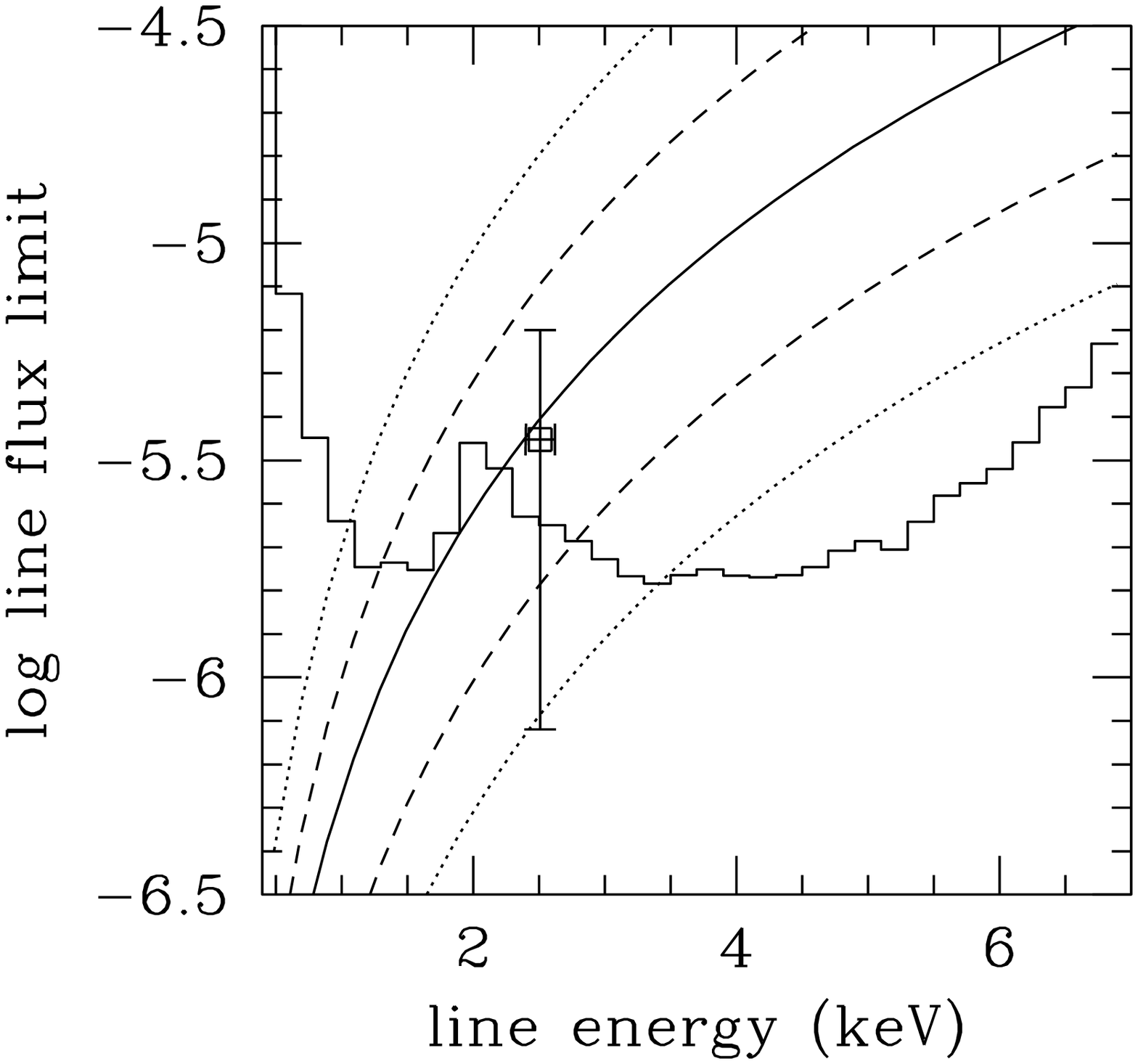}
\hfil
\caption{The histogram shows the $\Delta\chi^2=4.61$ upper limits for
  a line-free spectrum. The solid and broken lines show the expected
  line flux from sterile neutrino radiative decay in Willman 1,
  assuming $f_{\rm st}=1$ and taking into account the hadronic
  uncertainties in production (Asaka et al. 2007), which affect the
  relation between the mass and the mixing angle corresponding to a
  given value of $f_{\rm st}$. The dotted curves allows for a factor
  of two uncertainty in mass in either direction ($M_{7}=0.1-0.4$),
  {\it in addition} to the spread in possible production rate.}
\end{figure}

\subsection{Discussion}

\subsubsection{Comparison with Limits from Classical Milky Way Dwarf
Spheroidals}

We re-examine the {\it Suzaku} spectra of the Ursa Minor and Draco
``classical'' Milky Way dwarf spheroidals to check for consistency
with the line flux limits of the 2.5 keV candidate sterile neutrino
radiative decay emission line in Willman 1. The projected mass
estimate adopted in LKB within the {\it Suzkau} spectral source region
for Ursa Minor is $\sim 30\times$ our fiducial value for the mass
within the (smaller) Willman 1 {\it Chandra} extraction region. Draco
has a similar projected mass to that of Ursa Minor \citep{str07}.

Factoring in these mass ratios, and their respective distances, the
90\% limits of the Willman 1 line flux translates to a predicted flux
of $(3.2\pm 2.5)\times 10^{-5}\ {\rm photons}\ {\rm cm}^{-2}\ {\rm
s}^{-1}$ in Ursa Minor. In re-analyzing these spectra, we focus on the
1.3-5.2 keV energy region and, as in LKB, fit the {\it unsubtracted}
and unbinned spectrum to a model that includes power-law and Gaussian
non-X-ray background (PB) components (see LKB for details on how the
spectra are segmented, and other details). An additional line is not
required based upon simultaneous re-analysis of the {\it Suzaku} XIS1
(frontside-illuminated) and XIS03 (co-added backside illuminated)
spectra -- in fits where the line flux is tied in the two
spectra. However, fits to the XIS1 spectra, when analyzed
independently, are improved by the addition of such a line ($\Delta
C$=7, where C is the Cash-statistic; Cash 1979). The best fit line
energy is 2.41 keV with 90\% confidence limits 2.35-2.57 keV (and a
secondary local minimum at $\sim 2.5$ keV); the line flux is $(2.5\pm
1.9)\times 10^{-5}\ {\rm photons}\ {\rm cm}^{-2}\ {\rm s}^{-1}$ (Figure
10a) -- consistent with the expectations. However, the line is not
detected in the (less sensitive) XIS03 Ursa Minor spectrum: the upper
limit is $1.8\times 10^{-5}\ {\rm photons}\ {\rm cm}^{-2}\ {\rm
s}^{-1}$ (Figure 10b). The {\it Suzaku} spectra of Draco, reduced and
analyzed in an identical manner, do not require a 2.5 keV line. The
upper limit from simultaneous fits is $\sim 2.5\times 10^{-5}\ {\rm
photons}\ {\rm cm}^{-2}\ {\rm s}^{-1}$, with similar limits from the
individual detectors. Because of its larger distance, the expected
flux from Draco is $\sim 25$\% smaller than for Ursa Minor.

\begin{figure}
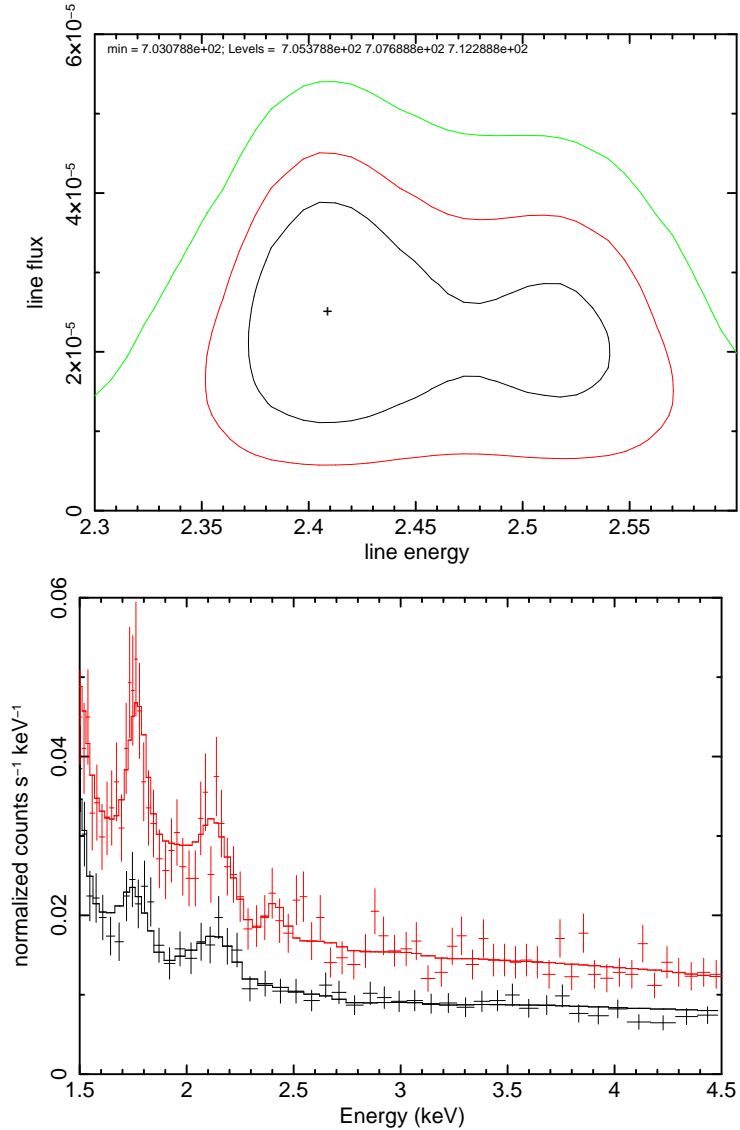

\centering
\includegraphics[scale=0.4,angle=270]{nfig10a.eps}
\hfil
\centering
\includegraphics[scale=0.4,angle=270]{nfig10b.eps}
\hfil
\caption{{\bf (a) top:} $\Delta C=2.3$, 4.61, and 9.21 confidence
contours for the emission line parameters, from the {\it Suzaku} XIS1
spectrum of Ursa Minor. {\bf (b) bottom:} XIS1 (red) and XIS03 (black)
spectra and best fit models that include a $\sim 2.4$ keV line (the
flux is $2.5\times 10^{-5}\ {\rm photons}\ {\rm cm}^{-2}\ {\rm
s}^{-1}$ in the former, and 0 in the latter).}
\end{figure}

\cite{sh09} analyzed archival data (ObsIDs 9589 and 9776) of the Draco
dwarf spheroidal galaxy. We have re-analyzed these data, essentially
replicating their reduction procedures, but deriving line flux upper
limits from the spectrum extracted from a $4'$ circular aperture by
application of the procedure used to derive limits for Willman 1
(Section 3). There are significant differences in the two
datasets. The Willman 1 observation is about three times deeper; and
was centered on the I-array instead of the S3 chip. The latter has
$\sim 30\%$ higher effective area at 2.5 keV,\footnote
{http://cxc.harvard.edu/proposer/POG/} but also a higher
background. Other factors that contribute to a lower background in the
Willman 1 observation are the lower point source detection threshold
(about 5 times as many point sources are detected), the higher
concentration of Willman 1 that allows us to use a a smaller spectral
extraction region, and the ability to pre-filter the particle
background (see section 2; the Draco observation was conducted in
FAINT, rather than VFAINT, data-mode). We derive a 90\% upper limit at
$\sim 2.5$ keV of $2.0\times 10^{-6}\ {\rm photons}\ {\rm
  cm}^{-2}\ {\rm s}^{-1}$ in Draco, consistent with the results of
\cite{sh09}. Based on an estimated projected mass of $\sim 6\times
10^6$ M$_{\odot}$ \citep{str07}, the expected flux is $\sim 75$\% that
in Willman 1 (Draco is twice as distant) -- {\it i.e.} $(2.5\pm
2)\times 10^{-6}\ {\rm photons}\ {\rm cm}^{-2}\ {\rm s}^{-1}$.

\subsubsection{Other Dark Matter Dominated Systems}

Some of the strongest claimed limits on sterile neutrino parameters
are derived from M31 \citep{wat06,birs08,briwrsh10}. These depend on
estimates of the dark matter content in regions where baryonic matter
is significant, if not dominant. While there is a fairly strong
consensus on the total mass in M31, there is a wide range of possible
mass decompositions (primarily) into stellar disk, stellar bulge, and
dark matter halo components \citep{ccf,cor09,saglia10} -- wider than
assumed in the papers referred to above. The best-fits presented in
\cite{cor09} are obtained using the \cite{burk95} dark matter density
distribution function. The resulting projected dark matter mass of
$2.9\times 10^{9}~M_\odot$ (for models with constrained virial mass)
in the 1.1-3 kpc annulus, demonstrates that a non-detection in M31
does not stand in contradiction to a feature in Willman 1 with
best-fit parameters we derive from the {\it Chandra} data
\citep{KL10}.

The diluting effects of X-ray emission associated with stellar and/or
hot plasma presents a potential complication in systems with prominent
baryonic components.  For galaxies, we estimate the contribution from
an old stellar population of mass $M_*$ to the luminosity emitted in
an energy interval $\Delta E$ at 2.5 keV, using the power-law indices
and normalizations in \cite{rev08}, as
\begin{equation}
L_{\Delta E}({\rm 2.5 keV})/M_*=1.0\times 10^{26}[1+40(1-f_{\rm
    res})]{{20\Delta E}\over E},
\end{equation}
where the ratio $\Delta E/E$ is normalized to a typical CCD spectral
resolution (at 2.5 keV) of 20.  The first term accounts for
cataclysmic variable and coronally active binary stars, as seen in
the Milky Way and M31 \citep{lw07,bg08}. The second accounts for low
mass X-ray binaries, with $f_{\rm res}$ the fraction of the LMXB flux
that is resolved out (which depends on the angular resolution and
exposure depth of the observation, as well as the LMXB luminosity
function). The ratio of this term to the sterile neutrino radiative
decay line emission from the dark matter mass $M_{\rm dark}$ is
\begin{equation}
[0.111+4.44(1-f_{\rm res})]{{20\Delta E}\over
  E}\Gamma_{-27}^{-1}{M_*\over {M_{\rm dark}}}.
\end{equation}
This emission, that may be augmented by discrete sources and hot gas
associated with a more recent star formation, may be non-negligible
where dark matter is subdominant or marginally dominant. 

In galaxy clusters and groups, and massive elliptical galaxies, that
are filled with hot ($\sim 1-10$ keV) plasma enriched with the
products of star formation, emission of a 2.5 keV sterile neutrino
decay line may be overwhelmed by $K\alpha$ emission lines at $\sim
2.45$ and $\sim 2.6$ keV from He-like and H-like S, respectively. For
example, in M87 the observed S XV line flux \citep{mat96} within $10'$
is $\sim 10\times$ that expected from the amount of dark matter
\citep{chu08} if it were composed of sterile neutrinos decaying at the
rate inferred in Willman 1.

Of course, the gravitational potential wells in dwarf spheroidals are
far too shallow (their observed velocity dispersions corresponding to
$\sim 10$ eV temperatures) to bind gas hot enough to collisionally
ionize S to this degree, and there are no significant sources of
photoionization. If hot plasma were somehow responsible for the
Willman 1 2.5 keV emission line, one would expect much stronger
accompanying emission in Si-K and Fe-L shell emission that is not
seen -- unless the elemental abundances are highly and uniquely
peculiar. When we attempt to fit the Willman 1 spectrum with a thermal
plasma model (best fit temperature, $kT\sim 2$ keV), we find that the
2.5 keV line may be accounted for only if the S/Fe abundance ratio
exceeds $50\times$ the solar ratio.

\subsubsection{Other Constraints}

The limits based on Lyman-$\alpha$ forest data \citep{blrv09} allow up
to 40\% of dark matter in the form of 5-keV sterile neutrinos produced
via Dodelson-Widrow (DW) scenario \citep{dw94}. However, if there is
any contribution from the Higgs boson decays, the resulting population
of relic sterile neutrinos has about three times smaller
free-streaming length than the DW population
\citep{k06,p08,pk08,boyan08b}.  The inclusion of the additional
``cold'' population of sterile neutrinos weakens or eliminates the
Ly-$\alpha$ bounds.  Although a detailed numerical analysis of such
mixed dark matter has yet to be performed, the approximate analytical
results indicate that 5~keV sterile neutrinos can account for 100\% of
dark matter in this case \citep{p08,boyan08a,boyan08b}.

\subsubsection{Instrumental Features in Proximity}

Improper subtraction of the PB could conceivably produce artefacts in
the subtracted spectrum that may be misinterpreted as a feature. We
address this issue, as follows, by applying an analysis procedure to
the Willman 1 spectrum along the lines that we implemented in
analyzing the {\it Suzaku} dwarf spheroidal spectra (see Section
4.2.1, LKB). We consider the full unsubtracted, and unbinned, spectrum
in the 2.2-5 keV band. The reduced-background PB spectrum (Section
2.4) is fit by a model consisting of a power-law and two Gaussian
emission lines with energies corresponding to the brightest Au
fluorescent features \citep{cha00}, and line-widths that are free to
vary. The addition of a line at 2.51 keV does not improve the fit, and
the $\Delta C=2.7$ line flux upper limit is $6\times 10^{-7}\ {\rm
  photons}\ {\rm cm}^{-2}\ {\rm s}^{-1}$. If we then fit the total
(unsubtracted) spectrum with the same baseline model, the addition of
a line at 2.51 keV improves the fit by $\Delta C=5.8$ and the best-fit
line flux of $3.5_{-2.5}^{+2.1}\times 10^{-6}{\rm photons}\ {\rm
  cm}^{-2}\ {\rm s}^{-1}$ ($\Delta C=2.7$ errors) is consistent with
the estimate based on the PB-subtracted spectrum. If we allow the line
energy to float, we find energy $2.57\pm 0.06$ keV and flux $4.2\pm
2.5\times 10^{-6}\ {\rm photons}\ {\rm cm}^{-2}\ {\rm s}^{-1}$
($\Delta C=4.6$ errors), again consistent with the initial
estimate. The improvement in fit over the baseline model is $\Delta
C=10.6$. If we then go back and refit the PB spectrum with an
additional line with energy fixed at 2.57 keV, we find that the
$\Delta C=2.7$ line flux upper limit is $1.6\times 10^{-6}\ {\rm
  photons}\ {\rm cm}^{-2}\ {\rm s}^{-1}$, and the inclusion of the
line does not improve the fit ($\Delta C=1.6$). These considerations
indicate that the 2.5 keV feature is not an artefact of inaccurate PB
subtraction. The unbinned PB and total spectra and best fit models are
shown in Figure 11ab; the same comparison with the data rebinned (in
the plot only) is shown along with the two best-fit models in Figure
12. Confidence contours are plotted in Figure 13.

\begin{figure}
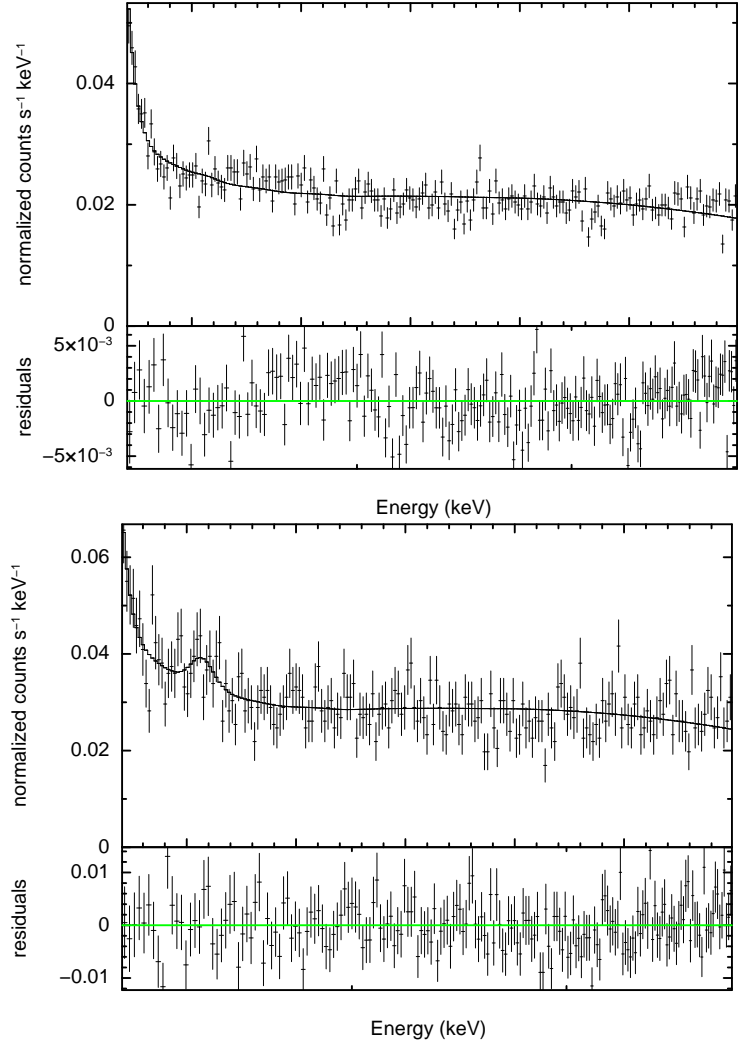

\centering
\includegraphics[scale=0.4,angle=270]{nfig11a.eps}
\hfil
\centering
\includegraphics[scale=0.4,angle=270]{nfig11b.eps}
\hfil
\caption{{\bf (a) top:} Unbinned PB spectrum and best fit model (see
  text). {\bf (b) bottom:} Same for total Willman 1
  spectrum. Inclusion of an additional emission line, with energy and
  flux consistent with that obtained from the PB-subtracted data,
  significantly improves the fit in the total, but not the PB,
  spectrum.}
\end{figure}

\begin{figure}
\centering
\includegraphics[scale=0.6,angle=270]{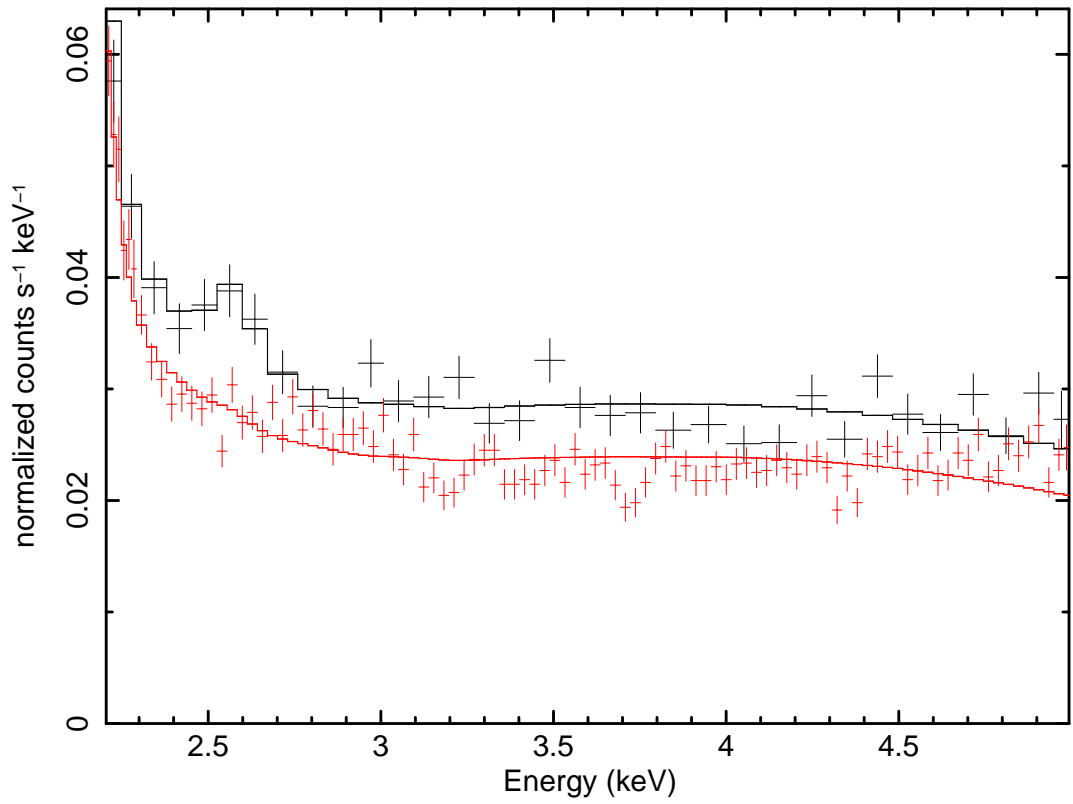}
\hfil
\caption{Spectra and models from Figure 11, rebinned and plotted
  together.}
\end{figure}

\begin{figure}
\centering
\includegraphics[scale=0.6,angle=-90]{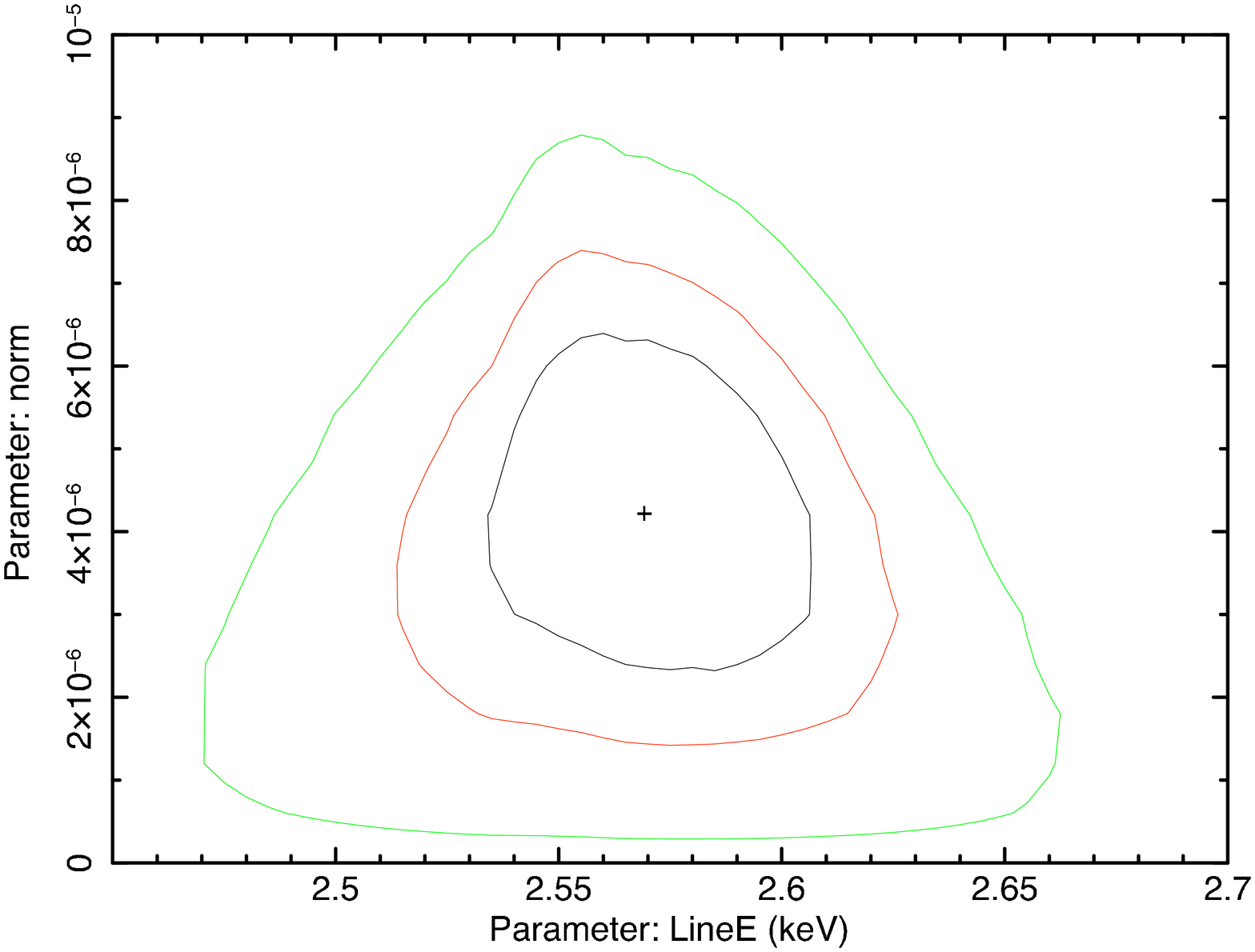}
\hfil
\caption{$\Delta C=2.3$, 4.61, and 9.21 confidence contours for the
  emission line parameters in fits to the unsubtracted, unbinned
  Willman 1 spectrum.}
\end{figure}

\subsubsection{Consistency, but not Corroboration}

Our re-examination of the other X-ray dwarf spheroidal datasets yields
an intriguingly similar weak 2.5 keV feature in the {\it Suzaku} XIS1
spectrum of Ursa Minor, but fails to confirm the presence of an
emission line at the expected strength in {\it Suzaku} and {\it
  Chandra} data on Draco and the {\it Suzaku} XIS03 spectrum of Ursa
Minor. \cite{briwrsh10} do not detect a feature in either the Sculptor
or Fornax dwarf spheroidals; although, it is notable that these
systems have relatively low mass-to-light ratios \citep{wol09} --
increasing the possibility of the presence of the diluting effects of
an unresolved stellar component (the former is known to have a
population of X-ray binaries; Maccarone et al. 2005) -- a factor that
is also an issue in M31 (see above). Detection of a 2.5 keV line from
sterile neutrinos is impeded in the most massive dark-matter dominated
systems, which are well known to retain hot metal-enriched
(interstellar, intragroup, or intracluster) gas with line-emission
from highly ionized S. Such sources of confusion are absent in the
extreme baryon-poor faint and ultra-faint dwarf spheroidals, making
them the cleanest targets for confirming or refuting the evidence for
5~keV sterile neutrino dark matter uncovered in Willman 1.

That is, the current absence of other positive detections may, in
every case, be accommodated with confirmation of the Willman 1
provisional detection reported here -- given the uncertainties in mass
profiles and various other factors discussed in this section. Our
measurement lies near the limit of faint source line fluxes attainable
with the best current detectors and, even in the most promising
targets, the possible presence of sterile neutrinos with the most
suitable parameters in terms of viability as a dark matter candidate
lie tantalizingly close to the edge of detectability. It is notable
that the Willman 1 X-ray observation is unique in that it represents
the only deep {\it Chandra} exposure of a very high mass-to-light
system, so that all sources of competing 2.5 keV line and continuum
emission of astrophysical origin are minimized. On the other hand, the
dynamical nature of Willman 1 remains unresolved and continues to be
investigated \citep{gut10,wil10}. Additional observations of this --
and/or other similar -- dwarf spheroidals may be required to settle
the question of whether the Willman 1 feature is real or spurious.

\section{Implications of a Detection}

\begin{figure}
\centering
\includegraphics[scale=0.8,angle=0]{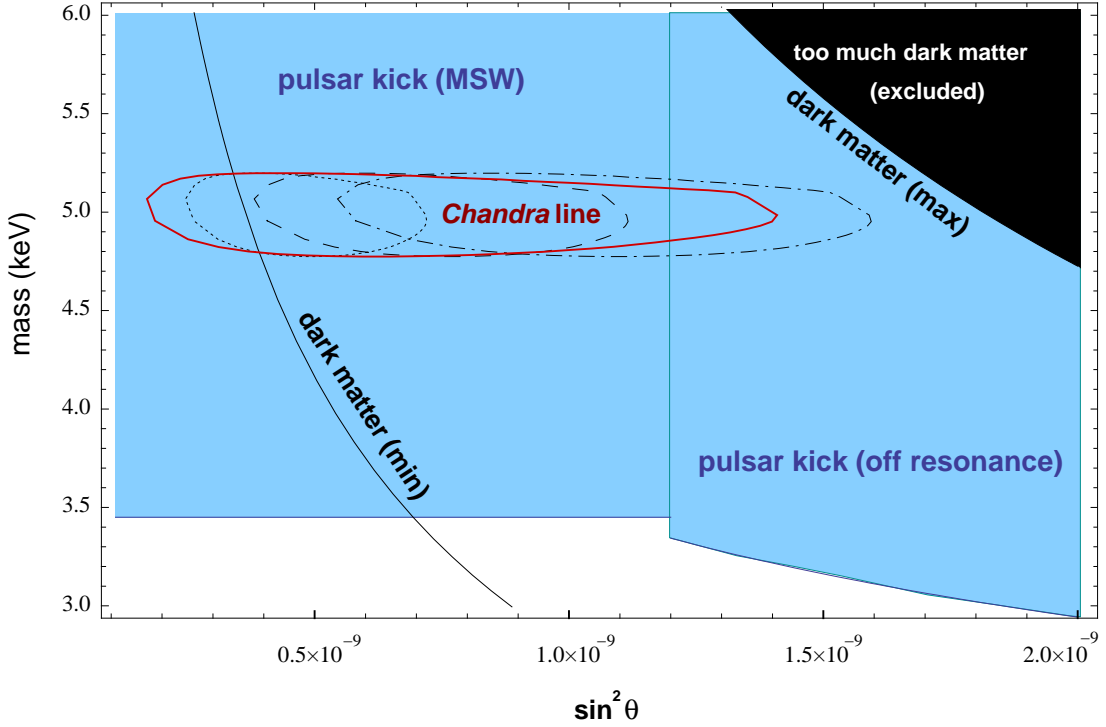}
\hfil
\caption{For sterile neutrinos that make up 100\% of dark matter, the
allowed range for the mass and the mixing angle inferred from the {\it
Chandra} data is shown by the solid contour. This range is consistent
with the region corresponding to production by non-resonant
oscillations alone that is circumscribed by the solid lines marked
``dark matter (min)'' and ``dark matter (max)''. If the primordial
abundance of sterile neutrinos is not assumed, the allowed range for
the mass and the mixing angle is shown by dotted, dashed, and
dot-dashed contours corresponding to maximum, average, and minimum
efficiency production by neutrino oscillations. A projected mass
within 55 pc of $2\times 10^6$ M$_{\odot}$ is assumed \citep{str08b}.
The inferred parameters are consistent with the pulsar kick mechanisms
based on either resonant (MSW) or non-resonant production in the
cooling neutron star (see text, Kusenko 2009).}
\label{fig:chandra_dm_pulsarkicks}
\end{figure}

One can take two different approaches to interpreting the spectral
line from a decaying relic particle. If one assumes dark matter is
solely composed of sterile neutrinos, then the number density of
particles is determined by the particle mass and the mixing angle is
inferred from the line flux. Alternatively, one can relax the
assumption equating the relic density and total dark matter density
\citep{hin09} -- e.g., there may be additional contributing dark
matter components. Regardless of any additional physics, sterile
neutrinos of a given mass and mixing angle are produced by neutrino
oscillations~\citep{dw94} at a rate that is calculable and cannot be
reduced by any unknown high-scale physics.\footnote{The only caveat to
this argument is the possibility of a non-standard cosmology with a
low reheat temperature, as pointed out by \cite{gelmini}.} The
contours for the allowed ranges of mass and mixing angle are somewhat
different depending on whether the relic sterile neutrino density is
assumed to be equal to the measured cosmological density of dark
matter~\citep{k06}. We show the results of both approaches in
Figure~\ref{fig:chandra_dm_pulsarkicks}.

Interpreted as a sterile neutrino radiative decay, the emission line
energy and 90\% flux limits imply $m_{\rm st}=5\pm 0.2$ keV,
$\Gamma_{\nu_s\rightarrow\gamma \nu_a}=(1.4 \pm 1.1)\times
10^{-27}(f_{\rm st}M_{7}/0.2)^{-1}$, and $\sin^2\theta=(7.8\pm 6.1)\times
10^{-10}(f_{\rm st}M_{7}/0.2)^{-1}$, where the projected mass in the
{\it Chandra} beam is normalized to the fiducial value of $2\times
10^6$ M$_{\odot}$. $\sin^2\theta=1.7\times 10^{-10}(M_{7}/0.2)^{-1}$
represents an absolute lower limit to the mixing angle, regardless of
the sterile neutrino production mechanism. If there was some DW
production, at least $9.5(M_{7}/0.2)^{-1}$\% of the dark matter must
have been produced in this manner (based on the minimum production and
lower line flux limit). If the DW mechanism is responsible for all of
the sterile neutrino production, $\sin^2\theta=(2.5-15.7)\times
10^{-10}(M_{7}/0.2)^{-1/2}$, where the lower (upper) limit corresponds
to the line flux lower (upper) limit and the maximum (minimum)
production; and, $f_{\rm st}>0.31(M_{7}/0.2)^{-1/2}$. The data is
consistent with DW production and $f_{\rm st}=1$ (Figure 9) for
$M_{7}=0.019-0.77$, where the lower (upper) limit is derived from the
minimum (maximum) flux and production. The corresponding mixing angle
for minimum (maximum) production is $\sin^2\theta=17.7(3.6)\times
10^{-10}$. In fact our best fit flux matches that expected for average
DW production and $f_{\rm st}=1$ for $M_{7}=0.18$ ($\sin^2\theta=8.2\times
10^{-10}$), which is close to our fiducial mass (Figure 9).

The allowed region may be further circumscribed if combined with
previously derived constraints. Limits have been derived using a
variety of datasets (mostly based on {\it Chandra} or {\it XMM-Newton}
data), employing a variety of data reduction, background mitigation,
and statistical techniques to derive upper limits, taking diverse
approaches (if any) to account for emission associated with baryons
(see above), and depend on different methods for estimates (and
associated uncertainties) of the dark matter content in the
field-of-view.  The limits based on {\it Chandra} deep field
measurements of the unresolved X-ray background originating in the
Milky way dark matter halo \citep{aba07} are perhaps the most directly
comparable in that deep {\it Chandra} ACIS-I data (some of
which was taken in VFAINT mode) was utilized. These constraints are
consistent with dark matter being composed of $m_{\rm st}=5$ keV
sterile neutrinos produced by the DW mechanism. {\it XMM-Newton}
observations of diffuse emission in the M31 halo are formally
inconsistent with our best-fit flux under the same
assumptions. However, taking the uncertainties into account, an ample
region of concordance is retained that includes parameters where
$f_{\rm st}=1$ and non-resonant oscillations are the dominant, or
sole, mechanism of sterile neutrino production (Section 4.2).

This interpretation has immediate implications for particle physics
and astrophysics. In particular, supernova physics is altered
dramatically by the emission of a 5~keV sterile neutrino that has a
mixing angle $\sin \theta\sim 10^{-5}-10^{-4}$. First, the emission
is anisotropic, with the direction of anisotropy set by the electron
spins polarized in the magnetic field~\citep{ks97,fkmp03}. While the
overall energy removed by sterile neutrinos with 5~keV mass and $\sin
\theta\sim 10^{-5}-10^{-4}$ is small, the anisotropy in the emission
of sterile neutrinos plays a central role in generating the supernova
asymmetries. Second, the lepton number, energy, and entropy transport
in the cooling neutron star are affected, and an increase in the
electron neutrino luminosity can preheat the material in front of the
shock wave, adding to the energy of the overall
explosion~\citep{hf07}. Third, the neutrino-driven kick enhances the
convection in front of the neutron star, which causes some energy
increase of the shock wave~\citep{fk06}. The numerical simulations of
this effect show the formation of asymmetric jets with the stronger
jet pointing in the direction of the pulsar motion~\citep{fk06}.

From the point of view of particle physics, the 5~keV sterile neutrino
with a small mixing angle can be accommodated in a seesaw mass matrix
that satisfies all the experimental and observational constraints
\citep{K09}. The values inferred from the \textit{Chandra}
observations allow for all dark matter to be produced via neutrino
oscillations, so that no additional production mechanism is
necessary. This, however, does not preclude the 5~keV mass from being
generated via the Higgs mechanism by a vacuum expectation value of an
electroweak-scale singlet Higgs boson, which may be discoverable by
upcoming experiments at the Large Hadron Collider~\citep{k06,pk08}.

The X-rays emitted by dark matter decay during the ``dark ages'' could
produce a strong enough ionization to enhance the formation of
molecular hydrogen, which, in turn, could speed up the formation of
the first stars~\citep{bk06}. The corresponding predictions for 21~cm
and other observations depend on the knowledge of the particle mass
and mixing angle~\citep{sbk07} that are greatly refined if the dark
matter particle has the mass and mixing inferred from \textit{Chandra}
observations of Willman 1.

Finally, there is no doubt that observational cosmology will take
advantage of our finding if it is confirmed. If dark matter emits a
narrow spectral line in X-rays, one can hope to map out the
three-dimensional dark matter distribution, including the redshift,
using future X-ray telescopes such as {\em
Astro-H}~\footnote{http://heasarc.gsfc.nasa.gov/docs/astroh/,
http://astro-h.isas.jaxa.jp/} and {\em
IXO}.\footnote{http://ixo.gsfc.nasa.gov/} Such data could be used to
study the expansion of the universe on cosmological distance scales.

\section{Summary and Concluding Remarks}

We presented results of a search for a radiatively decaying dark
matter emission line in the {\it Chandra} spectrum of the ultra-faint
dwarf spheroidal galaxy Willman 1. $\Delta\chi^2=9.2$ upper limits on
the line flux over the 0.4-7 keV bandpass were translated to an
allowed region in the sterile neutrino mass-mixing angle plane for the
case where all of the dark matter is composed of sterile neutrinos
regardless of how they are produced, and in the more general case
where their abundance is not assumed but is calculated under the
assumptions that minimize it. These constraints are consistent with
and therefore, given the uncertainties in mass determination,
reinforce those derived from previous X-ray analysis of dwarf
spheroidal galaxies (LKB, Riemer-Sorensen \& Hansen 2009).

We uncovered evidence of an emission line in the Willman 1 X-ray
spectrum with energy $2.51\pm 0.07 (0.11)$ keV and flux $[3.53\pm 1.95
  (2.77)]\times 10^{-6}\ {\rm photons}\ {\rm cm}^{-2}\ {\rm s}^{-1}$,
where the errors are 68\% (90\%) confidence limits. This was based on
an excess to the continuum measured by fitting the
particle-background-subtracted source spectrum, where the VFAINT
data-mode quiescent background reduction technique was
applied. Consistent parameters were found using maximum likelihood
fitting to the unbinned, unsubtracted spectrum. Re-examination of {\it
  Suzaku} data of the Ursa Minor dwarf spheroidal, and {\it Chandra}
data on the Draco dwarf spheroidal, reveal a possible confirmation in
the {\it Suzaku} XIS1 spectrum of Ursa Minor, but not in any of the
other spectra. Evidence of dark matter radiative decay in Willman 1
remains provisional due to the low level of significance, but
uncontradicted. In the short term, additional X-ray observations of
SDSS dwarf spheroidals are needed, while future high energy resolution
X-ray spectroscopy should prove definitive.

We derive the range of allowed sterile neutrino mass and mixing angle
implied by the 90\% confidence limits of line energy and flux,
assuming that either (1) dark matter solely consists of sterile
neutrinos regardless of how they are produced, or that (2) they are
solely produced by neutrino oscillations regardless of whether the
abundance matches that of dark matter as inferred from
cosmology. Remarkably, the two allowed regions strongly overlap --
that is, the observations are consistent with the narrow region in the
mass-mixing-angle plane where neutrino oscillations may produce all of
the dark matter (although it allows for some non-DW
production). Moreover, the inferred parameters are consistent with the
pulsar kick mechanisms based on either resonant or non-resonant
sterile neutrino production in cooling neutron stars. This unlikely
concordance intensifies the importance and urgency of additional
investigation.

\acknowledgments The authors thank James Bullock and Manoj Kaplinghat
for discussions of mass profiles, target selection, and other key
issues related to this work. We are also grateful to Beth Willman for
providing the latest information on the nature of Willman 1, the
referee for a thorough and constructive report, and members of the
astroparticle physics community too numerous to name for feedback via
e-mail and at scientific meetings. Support for this work was provided
by the National Aeronautics and Space Administration through {\it
  Chandra} Award Numbers G08-9091X and G09-0090X issued by the Chandra
X-ray Observatory Center, which is operated by the Smithsonian
Astrophysical Observatory for and on behalf of the National
Aeronautics Space Administration under contract NAS8-03060. The work
of AK was supported in part by DOE grant DE-FG03-91ER40662 and NASA
ATFP grant NNX08AL48G.


{}

\clearpage

\end{document}